\documentclass{appolb}
\usepackage{epsfig}

\usepackage{graphicx}

\usepackage{amssymb}
\usepackage[isolatin]{inputenc}
\usepackage{graphics}
\begin{document}
\title{Mass measurements of neutron-rich nuclei at JYFLTRAP
\thanks{Presented at XXX Mazurian lakes conference, Piaski, Poland.}%
}
\author{S. Rahaman\thanks{\emph{E-mail address:} saidur.rahaman@phys.jyu.fi}, V.-V. Elomaa, T. Eronen, U. Hager, J. Hakala, A.~Jokinen, A. Kankainen, 
J. Rissanen, C. Weber, J. Äystö and the IGISOL group 
\address{Department of Physics, P.O. Box 35 (YFL), FIN-40014 University of Jyväskylä, Finland}}
\maketitle
\begin{abstract}
The JYFLTRAP mass spectrometer was used to measure the masses of neutron-rich nuclei in the region of 28 $\leq$ \emph{N} $\leq$ 82 with uncertainties 
better than 10 keV. The impacts on nuclear structure and the r-process paths are reviewed. 
  
\end{abstract}
\PACS{ 21.10.Dr, 07.75.+h, 27.50.+e, 27.60.+j}
  
\section{Introduction}
Penning trap mass spectrometer is an important device for experimental physics \cite{SI06}. The confinement of ions in space by the combination of a 
homogeneous magnetic field and an electrostatic quadrupolar potential \cite{LB86} provides an ideal environment for performing mass measurements 
\cite{GB96} of nuclei far from the valley of stability. This offers a new opportunity for studying nuclear structure [4-7], stellar nucleosynthesis 
\cite{SR07,UH06a,AK06} and weak interaction tests \cite{TE06,TE06a} in nuclei. 

 The existence of nuclear shell structure as well as so called magic numbers have been experimentally verified among nuclei close to the valley of 
stability. However, it is natural to raise a question of whether the observed shell structure is influenced by the number of nucleons of opposite 
kind. A direct answer to this question remains open as masses of the most exotic nuclei are not yet known precisely.  In low mass region the vanishing 
magicity of the magic neutron numbers 20 and 28 has already been shown in ref \cite{NA91, OS93}. JYFLTRAP moves forward and has given considerable 
efforts over the years to measure the masses of neutron-rich nuclei in the region of 28 $\leq$ \emph{N} $\leq$ 82. These mass measurements have 
contributed to a better understanding of the neutron shell gap at \emph{N} = 40, 50, the proton shell gap at \emph{Z} = 28 and a large deformation 
around the \emph{N} = 59 region. These have been studied via the two-neutron separation and shell gap energies [4-8].
        
  Another obvious focus for mass measurements at JYFLTRAP is to provide accurate mass values for a better understanding of the nucleosynthesis 
pathways. Nuclear masses are one of the key input parameters for nucleosynthesis calculations. Along the rapid neutron capture process (r-process) 
paths neutron separation energies are rather low, typically 2 to 3 MeV \cite{JC91}. The precise masses and hence well defined neutron separation 
energies will help to pin down the location of the paths. Furthermore, in order to estimate the neutron capture rates, the masses and excitation 
energies should be known to better than 10 keV. Our precise mass results will substantially improve the r-process network calculations to minimize the 
models ambiguity.
 
 The combination of the JYFLTRAP \cite{VK04,AJ06} and the Ion Guide Isotope Separator On-Line (IGISOL) \cite{JA01} facility offers an unique 
opportunity to investigate neutron-rich nuclei including isotopes of refractory elements. The limiting factors are the half-lives and the production 
rates of the interesting nuclei. So far approximately 170 masses have been measured employing the JYFLTRAP setup. In this article nuclear structure 
effects revealed by these measurements are discussed. 

\section{The JYFLTRAP setup}
The experimental setup consists of a radiofrequency quadrupolar (RFQ) structure following by two cylindrical Penning traps inside a superconducting 
magnet of a field strength of 7 Tesla. The ions of interest were produced in proton-induced fission reactions by bombarding a thin (15 mg/cm$^{2}$) 
natural uranium target with a 30 MeV proton beam from K-130 cyclotron at the Department of Physics at the University of Jyväskylä. The continuous 30 
keV radioactive ion beam from IGISOL is first passed through a dipole magnet for isobaric mass selection and transported towards the RFQ for cooling 
and bunching. The bunched and cooled ion beam is then transferred to the first Penning trap for isobaric purification using the buffer gas cooling 
technique \cite{GS91} and finally a single ion sample is transferred to the second Penning trap for cyclotron frequency measurements using the 
time-of-flight ion cyclotron resonance technique \cite{MK95}. The mass of an ion of interest \emph{m} is then determined from the cyclotron frequency 
ratio \emph{r} of the well known reference ion ($\nu_{c, ref}$) to the ion of interest ($\nu_{c}$) using the formula
\begin{equation}
r = \frac{m - m_{e}}{m_{ref} - m_{e}} = \frac{\nu_{c, ref}}{\nu_{c}},
\end{equation}
where \emph{m} = mass of the interesting isotope, $m_{ref}$ =  mass of the known reference isotope, and $m_{e}$ is the electron mass. A detailed 
overview of JYFLTRAP will be presented in this issue in Ref. \cite{TE07}. 

\section{Discussion}
 
\begin{figure} \begin{center} 
\resizebox{0.99\textwidth}{!}{%
\includegraphics{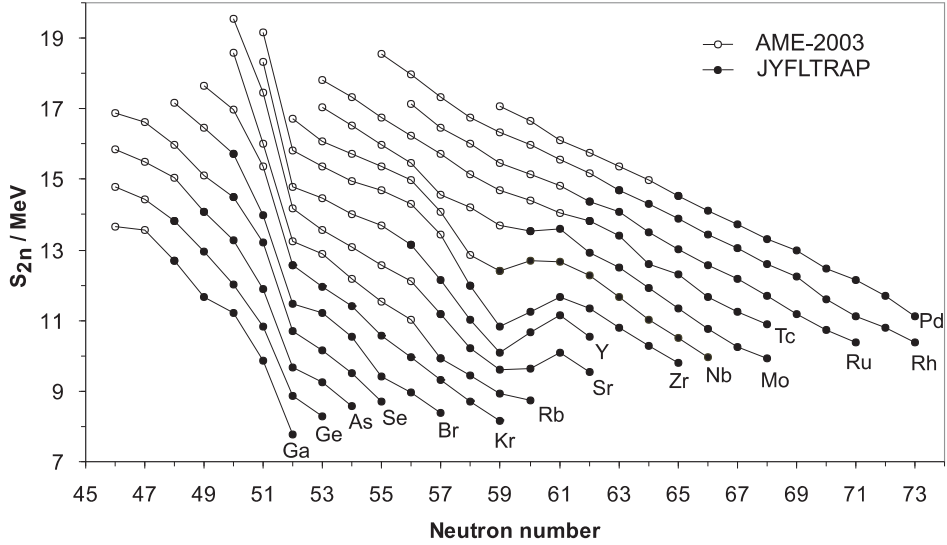}}
\caption{Experimental $S_{2n}$ energies as a function of neutron number for \emph{Z }= 31 - 46. The filled circles are from JYFLTRAP and the empty 
circles are from the AME03 mass table \cite{GA03}. Kr values indicated by the filled circles are from ref. \cite{PD07}. } \label{fig:1}
\end{center}\end{figure}

An up to date overview on the masses of neutron-rich nuclei investigated at JYFLTRAP is shown in Fig.~1 by plotting the two-neutron separation 
energies $S_{2n}$ as a function of the neutron number. $S_{2n}$ allows to observe nuclear structural changes along an isotopic chain. Generally, 
$S_{2n}$ values, decrease steadily as the neutron number increases, a characteristic of the liquid-drop model. This characteristic is well 
demonstrated for Ru, Rh and Pd isotopic chains in Fig.~1. The magic shell closure at \emph{N} = 50 can be seen as a sudden drop of the $S_{2n}$ values 
for all nuclei with \emph{N} = 51 and \emph{N} = 52.  

 Apart from these, a discontinuity appears in the $S_{2n}$ values between \emph{N} = 57 and 61 for isotopic chains from Rb to Tc. This discontinuity 
is strongest for the Zr isotopes and it reveals that the shell structure is undergoing a shape change from a spherical to a deformed shape. A more 
characteristic signature of this deformation in obtained by observing the change in the mean-square charge radii extracted from isotopic shifts 
measured in laser spectroscopy experiments. Fig.~2(A) shows the discontinuity in $S_{2n}$ and mean-square charge radii as a function of neutron number 
for Sr, Zr and Y isotopes. At a maximum discontinuity in $S_{2n}$ the mean-square charge radii increase suddenly. By observing the change in 
mean-square charge radii and subsequently in the quadrupole moments \cite{PL96} the Sr isotopes below \emph{N} = 59 were suggested to be weakly oblate 
and strongly prolate beyond \emph{N} = 59. Later similar sudden jumps in the mean-square charge radii at\emph{ N} = 60 were found in Zr and Y isotopes 
and a similar conclusion was drawn \cite{PC02,BC07}. 

\begin{figure} \begin{center} 
\resizebox{.999\textwidth}{!}{%
\includegraphics{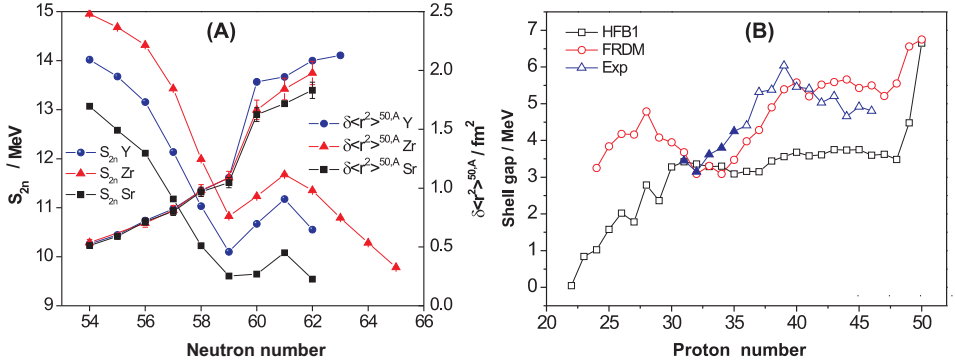}}
\caption{ \textbf{A}: $S_{2n}$ energies and mean-square charge radii as a function of neutron number for Sr, Zr and Y isotopes. \textbf{B}: 
Experimental neutron shell gap energies \cite{GA03} at \emph{N} = 50 compared with mass models \cite{JM06}. The filled triangles indicate the JYFLTRAP 
values.} \label{fig:2}
\end{center}\end{figure}

Shell gap quenching has been experimentally observed for \emph{N} = 20 \cite{NA91,OS93}. Some theoretical models \cite{RC99,MS02} predict a reduced 
neutron shell gap for \emph{N} = 50 towards lower \emph{Z} but no experimental data have been available until now. Figure~2(B) displays the 
experimental neutron shell gap for \emph{N} = 50 compared with two different models \cite{JM06}. A report from JYFLTRAP \cite{SR07} confirms the 
reduction of the shell gap energy towards $^{85}_{35}$Br. Later these measurements have been continued towards $^{81}_{31}$Ga \cite{JH07} where a 
further reduction in neutron shell gap energy was confirmed towards $^{82}_{32}$Ge. An increment in gap energy was observed for $^{81}_{31}$Ga 
compared to $^{82}_{32}$Ge which could be explained by the proton shell gap \emph{Z} = 28 approaching at $^{78}_{28}$Ni. FRDM model has predicted this 
effect clearly whereas HFB 1 deviates from the experimental observation.      

\begin{figure} \begin{center} 
\resizebox{0.999\textwidth}{!}{%
\includegraphics{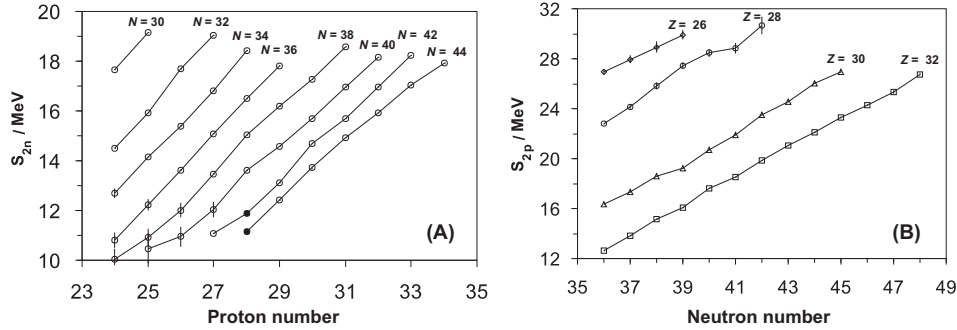}}
\caption{\textbf{A}: S$_{2n}$ plotted as a function of the proton number for even neutron chains. The filled circles are from JYFLTRAP and the empty 
circles are taken from ref. \cite{GA03}. A weak irregularity is observed for \emph{N} = 40 and 42 at \emph{Z} = 28. \textbf{B}: S$_{2p}$ plotted as a 
function of the neutron number for even proton chains. See text for details.} \label{fig:3}
\end{center}\end{figure} 
The \emph{N} = 40 subshell closure in $^{68}_{28}$Ni has been observed to be weaker than the \emph{Z} = 40 subshell closure in $^{90}_{40}$Zr.  
Experimental results are contradictory for the shell closure at $^{68}_{28}$Ni [30-33]. The first excited state in $^{68}_{28}$Ni is a 0$^{+}$ state 
observed at much higher energy than the first excited states in neighboring even-even nuclei. Also the 2$^{+}_{1}$ state in $^{68}_{28}$Ni has a large 
excitation energy and the $B(E2,0^{+}_{gs}\rightarrow 2^{+}_{1})$ value is quite small supporting a semidoubly-magic character of this nucleus 
\cite{OS02}. This conclusion, however, is not supported by recent JYFLTRAP measurement.  
  
Figure~3(A) shows $S_{2n}$ as a function of the proton number for even neutron numbers. The neutron shell gap energy is defined as $\Delta$(\emph{N}) 
= $S_{2n}(Z, N) - S_{2n}(Z, N+2)$ which is characterized by the vertical distance between two consecutive isotones in Fig.~3(A). A small enhancement 
of the neutron shell gap energy at \emph{N} = 40 for \emph{Z} = 28 is observed which can be explained by considering \emph{N} = 40 as a weak subshell 
closure \cite{SR07a}.  

The two-proton separation energies $S_{2p}$ are plotted as a function of the neutron number in Fig.~3(B) for even-\emph{Z} isotopic chains from 
\emph{Z} =26 to 32. In this plot the gap between the \emph{Z} = 28 and \emph{Z} = 30 chains gives the proton shell gap energy for \emph{Z} = 28. A 
tendency of decreasing proton shell gap energy is noticed for \emph{Z} = 28 at \emph{N} = 41. It is in agreement with the tensor force calculations 
which explained that with increasing number of neutrons (\emph{N} = 40 to 50) occupying the $\nu$1g$_{9/2}$ orbit, the $\pi$1f$_{7/2}$ and 
$\pi$1f$_{5/2}$ orbits come closer to each other \cite{TO06,TO05} resulting in a reduction of the proton shell gap energy. However, the next point at 
\emph{N} = 42 has a larger shell gap energy but it is based on extrapolated $^{68}$Fe mass. Therefore more experimental mass data are required, in 
particular precise masses of n-rich Fe isotopes to ascertain the observed trend.

\section{Conclusions and outlook}
More than 100 neutron-rich mass values in the vicinity of the astrophysical r-process paths have been determined at JYFLTRAP. A discontinuity in the 
$S_{2n}$ values has been observed around \emph{N} = 59 reflecting a change in the deformation. This effect was seen in \emph{Z} = 37 - 43 isotopic 
chains being strongest at \emph{Z} = 40. The \emph{N} = 50 neutron shell gap has been observed to decrease towards \emph{Z} = 32 and increased agian 
at \emph{Z} = 31. A small enhancement of the neutron shell gap energy has been noticed at $^{68}_{28}$Ni suggesting that there is a weak subshell 
closure. A reduction of the \emph{Z} = 28 proton shell gap energy is observed at \emph{N} = 41 which is in agreement with the tensor force prediction. 
However, more experimental masses are needed to study the trend towards \emph{N} = 50. JYFLTRAP will continue mass measurements of neutron-rich nuclei 
in order to understand the issues related to nuclear structure and astrophysics addressed in this paper.  

\section*{Acknowledgement}
This work has been supported by the TRAPSPEC Joint Research Activity project under the EU 6th Framework program "Integrating Infrastructure Initiative 
- Transnational Access", Contract Number: 506065 (EURONS) and by the Academy of Finland under the Finnish center of Excellence Program 2006-2011 
(Nuclear and Accelerator Based Physics Program at JYFL and project number 202256 and 111428).

\end{document}